# Brighter than the sun: Powerscape visualizations illustrate power needs in neuroscience and psychology

Pascal Wallisch
Department of Psychology
New York University

Statistical power is defined as the probability of detecting an effect that truly exists. Whereas statistical textbooks commonly implore researchers to aim for a power of at least 0.8 – in other words being able to detect a real effect 4 out of 5 times with the methods used in a given study – most actual research is performed without considering the concept of power at all (Téllez, García et al. 2015).

The effects of this neglect have been deleterious to the field.

One well-documented consequence is that much of research in neuroscience (Button, Ioannidis et al. 2013) and psychology (Chase and Chase 1976) has been severely underpowered, inadequate to find the effects the researchers were hoping for.

This is not a victimless sin. Lack of statistical power has actually more insidious consequences than "merely" missing some effects that truly exist. Running underpowered studies is also the royal road to false discovery. The reason being that lack of power leads to more broadly scattered estimates of the mean difference between groups, see figure 1 for an illustration of this effect.

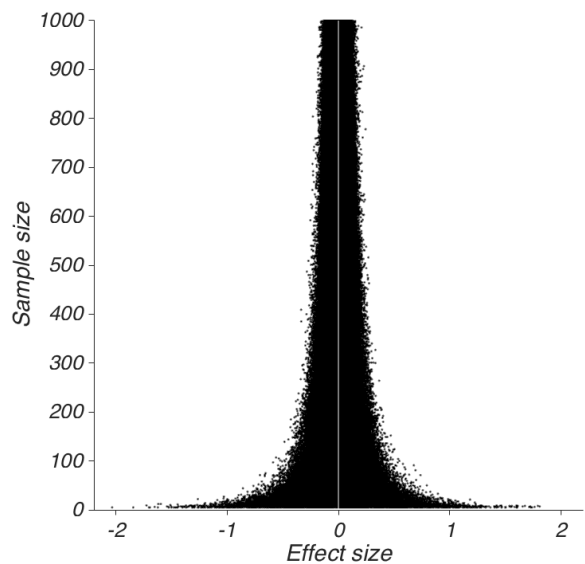

Figure 1: Funnel plot of surrogate data without a real effect. Each black dot represents a simulated study. Note that there are considerable numbers of studies reporting large (>0.5) effect sizes at low sample sizes, even if the effect size is zero and all data represent random noise.

As only positive (read significant) results are readily publishable, there are considerable incentives to run many underpowered studies and then report those that happen to scatter particularly widely. The technical term for this phenomenon is "*excess statistical significance*", effectively a reporting bias that demonstrably undermines the reproducibility of the field as a whole (Collaboration 2015).

This is particularly problematic in fields where real effect sizes can be expected to be small – and hard to distinguish from zero – such as in neuroscience and psychology. In practice, that seems to be exactly what is happening (Ioannidis 2005, Flint, Cuijpers et al. 2015). Note that this is not only a problem of obscure niche research. On the contrary, there is correlation between impact factor of the journal a paper is published in and the probability that the paper will have to be retracted (Brembs, Button et al. 2013), quite possibly due to the fact that power considerations are even rare for papers published in flagship journals such as Nature or Science (Tressoldi, Giofré et al. 2013). The fact that the problem seems to be most manifest in the most prominent journals that are read by the most people and have the most potential impact only amplifies these problems. It is important to note that although the lack of statistical power has been most intensively studied in neuroscience and psychology, there is emerging evidence that this problem is shared with much of modern science, including research on nutrition (Schoenfeld and Ioannidis 2013) and cancer (Tsilidis, Papatheodorou et al. 2012).

Curiously, this phenomenon has a long history. Instead of lacking evidence for statistical power, there has been compelling evidence of lacking power, going back for over half a century (Cohen 1962). Strangely, whereas this problem and its ramifications have been well known, it has resisted all attempts at fixing it. Statistical power in research studies remains alarmingly low and seems to be stubbornly resisting all efforts to raise it (Sedlmeier and Gigerenzer 1989, Rossi 1990).

Note that this has not been for lack of trying. Every so often, experts publish books on this very topic, providing – in principle – actionable advice on how to perform a power analysis before doing a study (Cohen 1977, Kraemer and Blasey 2015) .

So why are scientists not acting on this?

One consideration is that – thus far – scientists have been rewarded by the field for doing the wrong thing (Maxwell 2004). Running five underpowered small studies that show spurious effects is much more lucrative career-wise than running one adequately powered big one that shows that the theoretically desired effect doesn't actually exist.

In addition, there are also technical problems. One of them is that statistical power is hard to calculate analytically.
Statistical power depends on many things: The true effect size, the sample size, the particular statistical test used to analyze the data, the underlying population distribution, the reliability of the measurement methods and the significance threshold, among others.

Another issue concerns the fact that most researchers seem to have bad intuitions about how many participants – or units of analysis more generally – are needed to achieve sufficient statistical power if the hypothesized effect size can be expected to be small (Moher, Dulberg et al. 1994).

A possible remedy is to develop and provide tools that are easy for the non-expert to use and that yield results that are immediately intuitive.

This is what we attempt here – provide tools that can boost intuition for science practitioners.

Specifically we wrote a GUI in Matlab that allows users to see – at once glance – the rough power requirements of their proposed study, given a range of expected effect sizes.

We call this innovation "Powerscape", as it visualizes the expected power as a color as a function of both effect size and sample size, see figure 2.

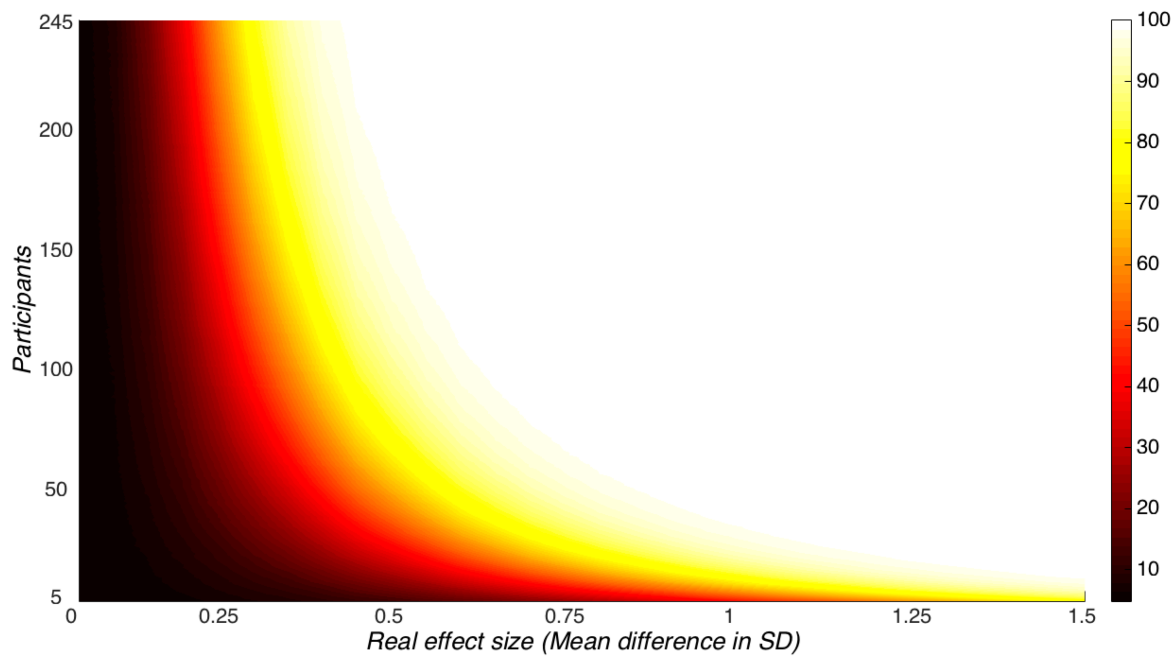

Figure 2: Powerscape for an independent samples t-test. Colors represent percentage of significant tests for a given effect and sample size, based on a simulation of 100,000 repeated test for each data point.

As we commonly aim for a power of at least 0.8, researchers would be well advised to look be in the yellow band above. To be on the left of the yellow band represents the underpowered range, to be in the white region represents overkill, which can be wasteful in terms of resources and also presents ethical concerns in cases where research participants and expected to endure pain and suffering in the study.

There is a different Powerscape for a range of several commonly used significance tests, such as t-tests for independent and paired samples, Wilcoxon's Rank sum test, etc. and the user can select each of these.

Users are then be able to pick an expected effect size, which will bring up the expected funnel plot (including confidence intervals) for that effect size, which allow them to pick an adequate sample size for their study. The program is freely available on [http://pensees.pascallisch.net/](http://pensees.pascallisch.net/)

Note that our simulations assume homogeneity of variances, the same number of participants in each of the two groups and that we are drawing from a normal distribution. In practice, this means that the powerscape represents the lower bound of the participants – or more generally units of analysis – in cases where we are concerned with neurons or the like. Real life demands will be higher to achieve a given power, as these assumptions are violated.

Interestingly, the powerscape visualization confirms our suspicion that intuitions about power needs can be misleading. There is a strong nonlinearity in terms of participant needs as a function of effect sizes. To achieve a given, high power, these needs are surprisingly low when the effect size is reasonably high, i.e. higher than 1 SD. This is consistent with everyday experience where dramatic group differences, such as the difference in height as a function of gender are obvious, even if the group difference is small. Conversely, the participant needs for effect sizes common in psychology and neuroscience – around 0.25 – are much higher than commonly assumed. Having 100 participants is not a large number and will yield power in the 0.4 range, which is about what we observe empirically in psychology (Cohen 1962, Rossi 1990). To have sufficient power to detect a typical effect in psychology will require on the order of about 500 study participants. To reliably detect small effects in complex and noisy environments, the corresponding number is well in the 1000s. Importantly, the Powerscape visualization makes this immediately obvious even to practitioners without a deeply sophisticated understanding of statistical nuances.

The reason we focused on sample size as the other variable – besides effect size – to put in the powerscape visualization is that researchers have limited control over most of the factors that influence power. True effect sizes and population distributions are given by nature. In addition, acceptable significance thresholds and the reliability of available methods are given by the field. It is the sample size and the type of significance test to use are the things that are most readily under the control of the researchers, and these are the things we let the user pick in our program.

An appreciation of power will only become more important in coming years, as there are increasing calls to make the significance threshold more conservative, as researchers are becoming ever more aware of problems like multiple comparison, selective reporting and researcher degrees of freedom in general (Simmons, Nelson et al. 2011, Gelman and Loken 2014, Simonsohn, Nelson et al. 2014, Franco, Malhotra et al. 2016). This development is well reasoned, but will further exacerbate power demands.

Luckily, there are technological solutions to many of these problems. For instance, environments like Psiturk (de Leeuw, Coenen et al.) allow researchers to easily gain data from thousands of participants online, allowing to reliably detect the presence of small effects buried in plenty of noise (Wallisch 2015).

But first, one needs to gain a deep appreciation for the need for power, even if one did not receive training in advanced statistical methods. We believe that intuition boosting tools like Powerscape are a step in that direction.